\documentclass[aps,prb,twocolumn,superscriptaddress,showpacs,floatfix]{revtex4}

\usepackage{bm}
\usepackage{amsmath}
\usepackage{amssymb}
\usepackage{graphicx}
\usepackage{color}
\usepackage{soul,xcolor}
\setstcolor{red}
\usepackage{ulem}
\usepackage{cancel}

\newcommand\Ccancel[2][red]

\usepackage{xcolor}\definecolor{ao}{rgb}{0.0,0.0,1.0}

\definecolor{br}{rgb}{1.0, 0.22, 0.0}

\usepackage{epsfig}

\newcommand{\up}{\uparrow}
\newcommand{\down}{\downarrow}

\def\sig{{\mbox{\boldmath{$\sigma$}}}}
\input{epsf}

\begin{document}
\title{Rashba proximity states in superconducting tunnel junctions}

\author{O. Entin-Wohlman}
\email{oraentin@bgu.ac.il}
\affiliation{Raymond and Beverly Sackler School of Physics and Astronomy, Tel Aviv University, Tel Aviv 69978, Israel}
\affiliation{Physics Department, Ben Gurion University, Beer Sheva 84105, Israel}

\author{R. I. Shekhter}
\affiliation{Department of Physics, University of Gothenburg, SE-412
96 G{\" o}teborg, Sweden}

\author{M. Jonson}
\affiliation{Department of Physics, University of Gothenburg, SE-412
96 G{\" o}teborg, Sweden}

\author{A. Aharony}
\affiliation{Raymond and Beverly Sackler School of Physics and Astronomy, Tel Aviv University, Tel Aviv 69978, Israel}
\affiliation{Physics Department, Ben Gurion University, Beer Sheva 84105, Israel}

\date{\today}

\begin{abstract}

We consider a new kind of superconducting proximity effect created by the tunneling  of ``spin split" Cooper pairs 
between two conventional superconductors connected by a normal conductor containing a quantum dot. 
The difference compared to the usual superconducting proximity effect is that the spin states of the tunneling Cooper pairs
are split into singlet and triplet components by the electron spin-orbit coupling, which is assumed to be active in the 
normal conductor only. We demonstrate that  the supercurrent carried by the spin-split Cooper pairs  can be manipulated  both 
mechanically and electrically  for  strengths of the  spin-orbit coupling  that can realistically be achieved by  electrostatic gates.

\end{abstract}
\pacs{72.25.Hg,72.25.Rb \vspace{2mm}
\\
Keywords: spin-orbit interaction, Rashba spin splitter, Josephson effect, electric weak link.}

\maketitle

\section{Introduction} 

\label{Intro}

The prominent role that the electronic spin plays in determining  the properties of solid-state devices has been at the forefront of experimental and theoretical research during the last decade. The topological surface electronic states,  that are formed due to a  strong spin-orbit coupling, with their vast potential for quantum computations [\onlinecite{Lehmann2007,Kloffel2013}] and spintronic applications [\onlinecite{Volbornik2011}] of spin-polarized currents,  are just a few conspicuous examples. Conducting nanostructures,  e.g.,  quantum dots, nanowires and nanorings, where the  mesoscopic behavior of the electrons is dominated by Coulomb correlations and quantum-phase coherence, are by now the tools of choice for studying spin-related phenomena, in particular effects induced by the spin-orbit coupling.  Composite mesoscopic structures comprising such nanometer-sized elements are currently of considerable interest  due to their applicability  in  quantum communication systems [\onlinecite{Hermelin2011,McNeil2011,Bertrand2016}]. The hope is to provide a coherent platform for flying qubits: moving two-state spinors, which may represent the electronic spin [\onlinecite{Popescu2004}] or any other pseudo-spin state, e.g.,  of particles moving in two coupled wires [\onlinecite{Yamamoto2014}].

Clearly, spin-state decoherece is detrimental to
spintronics applications involving, e.g., flying qubits. Reducing the scattering rate of spin-polarized electrons in order to preserve spin coherence is therefore
essential,  and is the reason why using superconducting materials have been considered. However, while electron transport in a superconductor is indeed fully coherent, the supercurrent carried by spin-singlet Cooper pairs in a conventional superconductor conducts charge but not spin.
If, on the other hand, the Cooper pairs could be spin polarized it would mean that a coherent, dissipationless spin current could be generated. Hence, it is highly desirable to find methods for generating spin-polarized Cooper pairs.
Recently, such a method --- involving the creation of
spin-polarized  Cooper pairs  in
superconducting weak links made of materials with a strong spin-orbit interaction (SOI) {--- was proposed. It was shown that the spin-structure of the Cooper pairs, injected into a non-superconducting material  in which the spin-orbit interaction is significant, can be
``predesigned"  in such a way that a net electronic spin-polarization is carried through an SOI-active  weak link that connects the superconducting leads.
The physics behind this phenomenon is the splitting of the transferred electronic states within the  weak link  with respect to
spin --  the so-called ``Rashba spin-splitting" [\onlinecite{Shekhter2017}]. As a consequence of this spin splitting, the electronic spin experiences quantum
fluctuations that lead to  a  ``triplet-channel" for  Cooper-pair transport through the link.

The ability to inject  electrons paired in a spin-triplet state into a conventional BCS superconductor from an SOI-active superconducting
weak link,  opens a route to all kinds of spintronics applications that can be implemented by using a dissipationless spin current.  However,  the appearance of spin-triplet Cooper pairs in a conventional BCS superconductor is a so-called proximity effect,  and spin-polarized Cooper pairs are 
present  only in the vicinity of the weak link.  In addition,  the triplet states are  vulnerable to any spin-relaxation mechanism in the
superconductor. A clever  composite device-design is therefore required to allow for the accumulation of paired electrons with a non-zero net spin,  while significantly blocking their spin relaxation.

In the present paper we suggest such a design,  and propose  a new type of a superconducting weak link in which
Rashba spin-split states involving pairs of time-reversed (``Cooper pair") states
can be established through the proximity effect. 
The generic component of the device is a quantum dot coupled to two  superconductors through 
SOI-active weak links in the form of  nanowires, as illustrated in Fig.~\ref{Fig1}. A significant 
advantage of the device is its relatively low spin relaxation rate -- a well-known property of  quantum dots [\onlinecite{Khaetskii2000,Yakushiji2004,Amasha2008,Hai2010,Rudner2010}].
The extent to which spin is accumulated on the dot can be controlled by electric fields that modify the SOI strength [\onlinecite{Nitta1997,Sato2001,Flensberg2010,Beukman2017}]. Another handle on the device is the possibility to tune mechanically the physical location of the dot, and thus affect the amount of tunneling between the two reservoirs [\onlinecite{Shekhter2013,Shekhter2014}].

\begin{figure}[htp]
\includegraphics[width=3.4in]{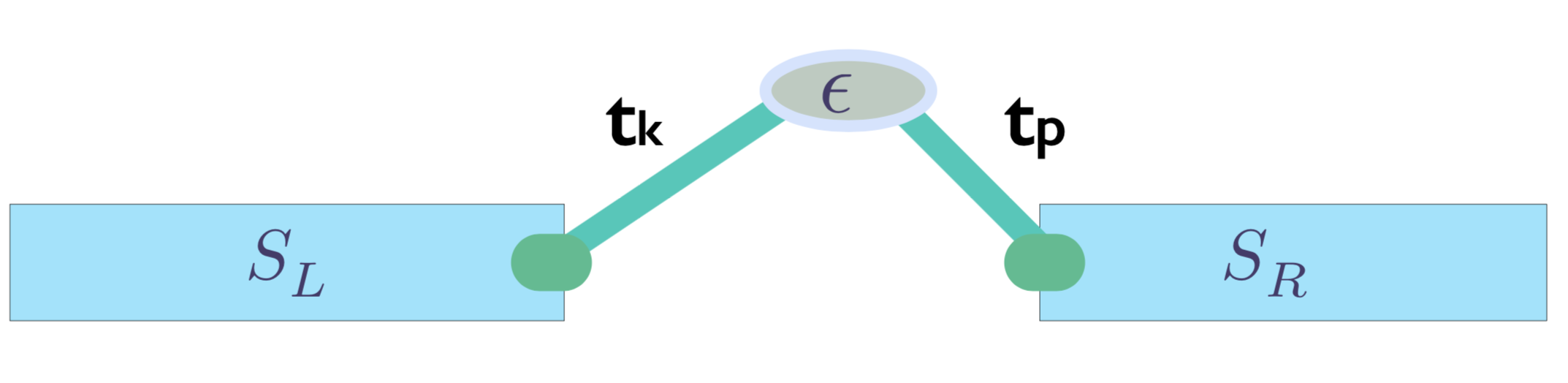}
\caption{Spin-orbit active superconducting weak-link. The quantum dot is represented by a single localized level, of energy $\epsilon$, and the superconducting reservoirs are denoted $S_{L}$ and $S_{R}$. The tunneling amplitude (a matrix in spin space) between the dot and the left (right) lead is denoted ${\bf t}_{\bf k}$ (${\bf t}_{\bf p}$).
}
\label{Fig1}
\end{figure}

The paper is organized as follows. Section \ref{Cooper} introduces
the Hamiltonian of our model and details the calculation of the transmission
of Cooper pairs between two superconductors connected by a  weak link on which the transferred electrons are subjected to a spin-orbit interaction. We include in the calculation two important effects. (i) The SOI on the left wire can differ from the one on the right wire (see Fig.~\ref{Fig1}), both in  strength and in the direction of the effective magnetic field that characterizes this interaction. (ii) The passage of a Cooper pair through the weak link can take place either by sequential tunneling, during which the two electrons tunnel one by one and the dot is at most singly occupied, or by  events in which the two electrons happen to reside simultaneously on the dot during the tunneling. In the latter case, one has to account for the Coulomb interaction on the dot. Obviously these two processes contribute disparately to the transport. We dwell on the two separate contributions and their dependence on the geometry of the junction in Sec. \ref{Results}.
Since we consider in Sec. \ref{Cooper} the transfer of Cooper pairs, in which the two electrons are in {\it time-reversed states}, 
we need to construct the relation between the (spin-dependent) tunneling amplitudes in these two states. This task is accomplished in Appendix \ref{TRtun}. The transmission of Cooper pairs is analyzed by studying the {\it equilibrium} Josephson current between the two reservoirs. In particular, we analyze the manner by which the spins of the tunneling electrons precess as they pass through the weak link. Technical details of this calculation are relegated to Appendix \ref{Pert}. 

Section \ref{Results} presents the results. 
We derive there explicit expressions for the spin-precession factor of each of the two processes alluded to above,  and explain the way  the disparity between the two reflects the coherence of the sequential single-electron tunneling process, and the incoherence of the double-electron one, during which the dot is doubly occupied.  We also analyze there the dependence of the Cooper pairs' transmission on the relative angle between the two directions of  the SOI's on the two wires. Our conclusions are discussed in Sec. \ref{Discussion}. 

\section{Tunneling of Cooper pairs}
\label{Cooper}

\subsection{Description of the model}
\label{Model}

The spin splitting of electrons that flow through a weak link in which the Rashba [\onlinecite{Rashba1960,Bychkov1984}] spin-orbit interaction is active,
can be understood within a semiclassical picture. As the electrons pass through the weak link, their spins precess around an effective magnetic field associated
with the 
SOI.   This spin dynamics  splits the electron wave function into different
spin states and yields a certain probability, which can be controlled externally, for the spins to be flipped as they
emerge from the weak link [\onlinecite{Shekhter2016}].
The 
spin-splitting phenomenon becomes more complicated when 
the transmission of {\it a pair of electrons in two time-reversed states} through an SOI-active weak link is considered. In that case there are two types of tunneling events, those where the electrons are transferred one by one sequentially, and those in  which the dot is doubly occupied during the tunneling. 
This is taken into account  in our model by representing  the  quantum dot  as a single localized level of energy $\epsilon$ which can accommodate two electrons in  two spin states (``up" and ``down"). In our simple model the reservoirs that supply the electrons are two bulk BCS superconductors; these are coupled together by a nanowire on which the quantum dot is located. When on the dot,   the spin state of one electron of the Cooper pair is projected on the spin-up state  of the dot,  and that of the other on the spin-down state [\onlinecite{axis}].   We find  that the projection breaks the coherent evolution of the spin states. The Pauli principle is assumed to be effective only on the quantum dot; elsewhere the passage of the electrons in and out of the dot is viewed as a single-electron tunneling event, whose amplitude includes the electronic spin precession [\onlinecite{pauli}].

The Hamiltonian of the  junction illustrated in Fig.~\ref{Fig1} reads
\begin{align}
{\cal H}={\cal H}^{}_{0}+{\cal H}^{}_{\rm tun}\ .
\end{align}
The Hamiltonian ${\cal H}_{0}$ pertains to the decoupled system, and includes the Hamiltonian of the quantum dot and that of the superconducting leads,
\begin{align}
{\cal H}^{}_{0}=
\sum_{\sigma}\epsilon d^{\dagger}_{\sigma}d^{}_{\sigma}+U
d^{\dagger}_{\up}d^{}_{\up}d^{\dagger}_{\down}d^{}_{\down}
+\sum_{\alpha=L,R}{\cal H}^{\alpha}_{\rm lead}\ .
\label{H0}
\end{align}
The operator $d^{}_{\sigma}$ ($d^{\dagger}_{\sigma}$) annihilates (creates) an electron in the spin state $|\sigma\rangle $ on the dot, and $U$ denotes the Coulomb repulsion. 
The BCS  leads
are described  by the annihilation (creation) operators  of the electrons there, $c^{}_{{\bf k}({\bf p})\sigma}$ ($c^{\dagger}_{{\bf k}({\bf p})\sigma}$).
[${\bf k}$ (${\bf p}$) enumerates the single-particle orbital states  on the left (right) lead.] Denoting by  $
\epsilon^{}_{k(p)}$  the single-electron energy measured relative to the common chemical potential of the device, 
the Hamiltonian of the leads is
\begin{align}
&{\cal H}^{\alpha=L(R)}_{\rm lead}=\sum_{{\bf k} ({\bf p}),\sigma}\epsilon^{}_{ k(p)}c^{\dagger}_{{\bf k}({\bf p})\sigma}c^{}_{{\bf k}({\bf p})\sigma}\nonumber\\
&-\Delta^{}_{L(R)}
\sum
_{{\bf k}({\bf p})}(e^{i\phi^{}_{L(R)}}c^{\dagger}_{{\bf k}({\bf p})\up}c^{\dagger}_{-{\bf k}(-{\bf p})\down}+{\rm H.c.})\ ,
\label{Hl}
\end{align}
where
$\Delta_{L(R)}$ and $\phi_{L(R)}$ are
the amplitude and the phase of the superconducting order parameters.

The tunneling Hamiltonian 
is the key component of our model, 
\begin{align}
{\cal H}^{}_{\rm tun}={\cal H}^{}_{LD}+{\cal H}^{}_{RD}+{\rm H.c.}\ ,
\end{align}
where
\begin{align}
{\cal H}^{}_{L(R)D}=
\sum_{{\bf k}({\bf p}),\sigma,\sigma'}c^{\dagger}_{{\bf k}({\bf p})\sigma}[{\bf t}^{L(R)D}_{{\bf k}({\bf p})}]^{}_{\sigma\sigma'}d^{}_{\sigma'}
\ .
\label{Htun}
\end{align}
The probability amplitude for the transfer of an electron from the spin state $|\sigma'\rangle$ on the  dot to the state $|{\bf k}({\bf p}),\sigma\rangle$ in the left (right) reservoir
is 
$[{\bf t}^{L(R)D}_{{\bf k}({\bf p}) }]_{\sigma\sigma'}$, which allows for  spin flips during the tunneling.
This amplitude is conveniently separated
into a (scalar) orbital amplitude, and a unitary matrix (in spin space), denoted ${\bf W}$,  that contains the
effects of the SOI (whether
of the Rashba [\onlinecite{Rashba1960,Bychkov1984}] or the Dresselhaus [\onlinecite{Dresselhaus1955}] type), and also 
the dependence on the
spatial direction of the SOI-active wire. The spin-orbit interaction associated with strains is briefly mentioned in Sec. \ref{Results}.
For
the linear  SOI [\onlinecite{Shekhter2017}], the tunneling amplitude can be presented in the form
\begin{align}
{\bf t}^{L(R)D}_{\bf k}=it^{}_{L(R)}e^{-ik^{}_{\rm F}d^{}_{L(R)}}{\bf W}^{L(R)D}_{}\ ,
\label{GF}
\end{align}
where $k_{\rm F}$ is the Fermi wave vector in the leads,  and  $d_{L(R)}$
is the length of the bond between
the left (right) lead and the dot. 
The generic form of ${\bf W}$ is
[\onlinecite{Shahbazyan1994,Ora2005}]
\begin{align}
{\bf W}^{L(R)D}_{}=a^{}_{L(R)}+i\sig\cdot{\bf b}^{}_{L(R)}\ ,
\label{W}
\end{align}
where $a_{L(R)}$ is a real scalar and ${\bf b}_{L(R)}$ is a real vector (determined by the symmetry axis of the SOI and the spatial direction of the weak link), with
$a^{2}_{L(R)}+{\bf b}^{}_{L(R)}\cdot{\bf b}^{}_{L(R)}=1$ ($\sig$ is the vector of the Pauli matrices).
In the absence of the spin-orbit interaction
${\bf W}^{L(R)}$ is just the unit matrix, namely, $a_{L(R)}=1$ and ${\bf b}_{L(R)}=0$\ .
Explicit expressions for the spin-orbit coupling are discussed in Sec. \ref{Results}.

Since the two electrons of a Cooper pair are in two states connected by the time-reversal transformation,  one has to consider also the tunneling amplitude between the  time-reversed states of $|\sigma'\rangle $ and $|{\bf k}({\bf p}),\sigma\rangle$, which we denote by an overline, i.e.,  
$|\overline{\sigma}'\rangle\equiv (i\sigma_y)\vert\sigma'\rangle$ and $\vert -{\bf k}(-{\bf p}),\overline{\sigma}\rangle$. The corresponding  amplitude, 
$[\overline{\bf t}^{L(R)D
}_{{\bf k}({\bf p})}]_{\overline{\sigma}\overline{\sigma}'}$, describes the
 transfer of an electron from
the  spin state $\vert\overline{\sigma}'\rangle $ on the dot to the state
$\vert -{\bf k}(-{\bf p}),\overline{\sigma}\rangle$ in the left (right) lead.
It is given by
\begin{align}
\overline{\bf t}^{L(R)D
}_{{\bf k}({\bf p})} \equiv  \hat{\bf  T} {\bf t}^{L(R)D
}_{{\bf k}({\bf p})} \hat{\bf T}^{-1}_{}
\,; \quad \hat{\bf  T}=K (i\sigma^{}_y)\ ,
\label{tbar}
\end{align}
($\sigma_y$ is a Pauli matrix,  and $K$ is the complex conjugation operator). We derive in Appendix \ref{TRtun}
the relation
\begin{align}
[\overline{\bf t}^{L(R)D
}_{{\bf k}({\bf p})}]^{}_{\overline{\sigma}\overline{\sigma}'}=[({\bf t}^{L(R)D
}_{{\bf k}({\bf p})})^{\ast}_{}]^{}_{\sigma\sigma'}\ ,
\label{IMR}
\end{align}
which is of paramount importance in our considerations.


\subsection{The particle current}
\label{Current}


The flow of electrons between the two superconductors
is analyzed by studying the equilibrium Josephson current, i.e.,  the rate by which electrons leave the left (or the right)  superconductor, when  the chemical potentials of the two reservoirs are identical (and therefore a flow of quasi-particles is prohibited by the  gap in the quasiparticle density of states in the superconducting leads).

Using units in which $\hbar=1$, this rate is
\begin{align}
J^{}_{}=-(d/dt)\langle\sum_{{\bf  k},\sigma}
c^{\dagger}_{{\bf k}\sigma}c^{}_{{\bf k}\sigma}\rangle=-2\,{\rm Im}\langle {\cal H}^{}_{LD}(t)\rangle\ ,
\label{JL}
\end{align}
where the angular brackets denote quantum averaging.
It is evaluated
using the $S-$matrix, 
\begin{align}
\langle {\cal H}^{}_{LD}(t)\rangle = \langle S^{-1}(t,-\infty) {\cal H}^{}_{LD}(t) S(t, -\infty) \rangle\ ,
\end{align}
with ${\cal H}_{LD}(t) = \exp[i{\cal H}_{0}t] {\cal H}_{LD} \exp [-i{\cal H}_{0}t]$, and  the quantum average is
 with respect to ${\cal H}_0$.
As the leading-order contribution to $J$ is fourth-order in the tunneling Hamiltonian,
it is found from the expansion up to  third order of the $S$-matrix  [\onlinecite{Gisselfalt1996}],
\begin{align}
\label{S}
&S(t, -\infty) = 1-i \int_{-\infty}^t dt_1 {\cal H}_{\rm tun}(t_1)\nonumber\\
&- \int_{-\infty}^t dt^{}_1  \int_{-\infty}^{t^{}_{1}} dt^{}_2 {\cal H}_{\rm tun}(t^{}_1) {\cal H}_{\rm tun}(t^{}_2) \\ 
&+i   \int_{-\infty}^t dt^{}_1  \int_{-\infty}^{t^{}_1} dt^{}_2 \int_{-\infty}^{t^{}_2} d t^{}_3 {\cal H}^{}_{\rm tun}(t^{}_1) {\cal H}^{}_{\rm tun}(t^{}_2) {\cal H}_{\rm tun}(t^{}_3)\ .\nonumber
\end{align}
The energy level on the dot is assumed to lie well above the chemical potential 
of the leads,  and thus the small parameter of the expansion is $\Gamma/\epsilon$, where
$\Gamma=\Gamma_{L}^{}+\Gamma^{}_{R}$
is the total width of the resonance level created on the dot due to the coupling with the bulk reservoirs (each coupling induces a  partial width $\Gamma_{L(R)}$, see Appendix \ref{Pert}). This implies that the perturbation expansion is carried out on a dot which is  empty when decoupled [\onlinecite{occupied.dot}].
We list and discuss the relevant terms in the expansion of $J$ in Appendix \ref{Pert}. 
 
As mentioned, the total Josephson current in our junction is due to the  two processes available for the  tunneling pairs.
In the first,  whose contribution is denoted $J^{\rm s}$, double occupancy on the dot does not occur, and the transfer of the electron pair is accomplished
by a sequential tunneling of the paired electrons one by one.
In the second process that contributes $J^{\rm d}$, the dot is doubly occupied during the tunneling.
The (lengthy but straightforward) calculation presented in Appendix \ref{Pert} yields
\begin{align}
J=J^{\rm s}_{}+J^{\rm d}_{}\ ,
\label{Jtot}
\end{align}
where
\begin{align}
J^{\rm s}_{}&=I^{}_{0}F^{\rm s}(\epsilon/\Delta){\cal A}^{\rm s}_{}\ ,\nonumber\\
J^{\rm d}_{}&=2I^{}_{0}F^{\rm d}(\epsilon/\Delta,U/\Delta){\cal A}^{\rm d}_{}\ .
\label{Jsd}
\end{align}
These results are obtained, for simplicity, in the zero-temperature limit, and for $\Delta_{L}=
\Delta_{R}\equiv\Delta$.
The common factor $I_{0}$ in the  expressions for $J^{\rm s}$ and $J^{\rm d}$
is the Josephson amplitude of the interface between the two superconductors (i.e., when the localized level on the dot as well as the SOI are ignored), $I_{0}=
2\sin(\phi_{R}-\phi_{L})[\Gamma_{L}\Gamma_{R}/\Delta]$.
The disparity of the two tunneling processes comes into play in the other two factors  in Eqs. (\ref{Jsd}). 
The functions $F^{\rm s}$ and $F^{\rm d}$ [\onlinecite{Glazman1989}], given in Eqs. (\ref{Fs}) and (\ref{Fd})
and reproduced here for convenience, 
\begin{align}
&F^{\rm s}_{}(\widetilde{\epsilon})=
\int_{-\infty}^{\infty}\frac{d\zeta^{}_{k}}{\pi}
\int_{-\infty}^{\infty}\frac{d\zeta^{}_{p}}{\pi}
\frac{1}{{\rm cosh}\zeta^{}_{k}+\widetilde{\epsilon}}\nonumber\\
&\times
\frac{1}{{\rm cosh}\zeta^{}_{k}+{\rm cosh}\zeta^{}_{p}}\frac{1}
{{\rm cosh}\zeta^{}_{p}+\widetilde{\epsilon}}\ ,\ \ \widetilde{\epsilon}=\frac{\epsilon}{\Delta}\ ,
\label{Fsm}
\end{align}
and
\begin{align}
&F^{\rm d}_{}(\widetilde{\epsilon},\widetilde{U})=
\int_{-\infty}^{\infty}\frac{d\zeta^{}_{k}}{\pi}
\int_{-\infty}^{\infty}\frac{d\zeta^{}_{p}}{\pi}
\frac{1}{{\rm cosh}\zeta^{}_{k}+\widetilde{\epsilon}}\nonumber\\
&\times
\frac{1}{2\widetilde{\epsilon}+\widetilde{U}}\frac{1}{
{\rm cosh}\zeta^{}_{p}+\widetilde{\epsilon}}\ ,\ \ \widetilde{\epsilon}=\frac{\epsilon}{\Delta}\ ,\ \ \widetilde{U}=\frac{U}{\Delta}\ ,
\label{Fdm}
\end{align}
convey the effect of the resonance on the dot, and the Coulomb repulsion there. As seen, the Pauli principle tends to reduce the contribution
of the second tunneling process.
The last factors, ${\cal A}^{\rm s}$ and ${\cal A}^{\rm d}$, describe the amount of spin precession. Their detailed analysis is the topic of the next subsection.
These two factors become 1 in the absence of the SOI.
It follows that in the absence of the SOI, the Josephson current of our junction, $J_{0}$, is
\begin{align}
J^{}_{0}=I^{}_{0}[F^{\rm s}_{}(\epsilon/\Delta)+2F^{\rm d}(\epsilon/\Delta,U/\Delta)]\ .
\end{align}
The normalized Josephson current, i.e., $J$ of Eq. (\ref{Jtot}) divided by $J_{0}$, is
\begin{align}
\frac{J}{J^{}_{0}}=\frac{F^{\rm s}_{}(\epsilon/\Delta){\cal A}^{\rm s}_{}+2F^{\rm d}_{}(\epsilon/\Delta,U/\Delta){\cal A}^{\rm d}_{}}{F^{\rm s}_{}(\epsilon/\Delta)+2F^{\rm d}(\epsilon/\Delta,U/\Delta)} \ .
\label{NorJ}
\end{align}


\subsection{Spin precession}
\label{SP}


In the sequential tunneling process, where the pair of electrons tunnel one by one, their spin-precession factor  is
\begin{align}
&{\cal A}^{\rm s}_{}=\frac{1}{2}\sum_{\sigma,\sigma'}\sum_{\sigma^{}_{L},\sigma^{}_{R}}{\rm sgn}(\sigma^{}_{L}){\rm sgn}(\sigma^{}_{R})\nonumber\\
&\times [({\bf W}^{LD}_{})^{\ast}_{}]^{}_{\sigma^{}_{L}\sigma'}
[{\bf W}^{LD}_{}]^{}_{\sigma^{}_{L}\sigma}
[({\bf W}^{DR}_{})^{\ast}_{}]^{}_{\sigma'\sigma^{}_{R}}
[{\bf W}^{DR}_{}]^{}_{\sigma\sigma^{}_{R}}\ ,
\label{Asm}
\end{align}
while the spin-precession factor of the tunneling process during which the dot is doubly occupied is 
\begin{align}
&{\cal A}^{\rm d}_{}=\frac{1}{2}\sum_{\sigma}\sum_{\sigma^{}_{L},\sigma^{}_{R}}{\rm sgn}(\sigma^{}_{L}){\rm sgn}(\sigma^{}_{R})\nonumber\\
&\times
[({\bf W}^{LD}_{})^{\ast}_{}]^{}_{\sigma^{}_{L}\sigma}
[{\bf W}^{LD}_{}]^{}_{\sigma^{}_{L}\sigma}
[({\bf W}^{DR}_{})^{\ast}_{}]^{}_{\sigma\sigma^{}_{R}}
[{\bf W}^{DR}_{}]^{}_{\sigma\sigma^{}_{R}}\ .
\label{Adm}
\end{align}
[See the derivation in Appendix \ref{Pert} that leads to Eqs. (\ref{As}) and (\ref{Ad}), reproduced in Eqs. (\ref{Asm}) and (\ref{Adm}) for convenience.]

It is illuminating to scrutinize the summations over the spin indices. The factor ${\cal A}^{\rm s}$ can be written as
\begin{align}
{\cal A}^{\rm s}_{}&=\frac{1}{2}
\sum_{\sigma^{}_{L},\sigma^{}_{R}}{\rm sgn}(\sigma^{}_{L}){\rm sgn}(\sigma^{}_{R}) [({\bf W}^{LR}_{})^{\ast}_{}]^{}_{\sigma^{}_{L}\sigma^{}_{R}}
[{\bf W}^{LR}_{}]^{}_{\sigma^{}_{L}\sigma^{}_{R}}\nonumber\\
&=
|[{\bf W}^{LR}_{}]^{}_{\up\up}|^{2}_{}-|[{\bf W}^{LR}_{}]^{}_{\up\down}|^{2}_{}
\ ,
\label{Asm1}
\end{align}
where the direct spin part of the tunneling amplitude from the right lead to left one is
\begin{align}
[{\bf W}^{LR}_{}]^{}_{\sigma^{}_{L}\sigma^{}_{R}}
=\sum_{\sigma}
[{\bf W}^{LD}_{}]^{}_{\sigma^{}_{L}\sigma}
[{\bf W}^{DR}_{}]^{}_{\sigma\sigma^{}_{R}}\ .
\label{Alr}
\end{align}
In contrast, the spin-precession factor of the double-occupancy channel cannot be expressed in terms of the direct amplitudes;
we find
\begin{align}
{\cal A}^{\rm d}_{}&=
(|[{\bf W}^{LD}_{}]^{}_{\up\up}|^{2}_{}
-|[{\bf W}^{LD}_{}]^{}_{\up\down}|^{2}_{})\nonumber\\
&\times(|[{\bf W}^{DR}_{}]^{}_{\up\up}|^{2}_{}
-|[{\bf W}^{DR}_{}]^{}_{\up\down}|^{2}_{})\ .
\label{adm}
\end{align}
One notes that ${\cal A}^{\rm d}$, Eq. (\ref{adm}),  is a product of two factors of the same structure as the single factor in ${\cal A}^{\rm s}$, Eq. (\ref{Asm1}). Our interpretation is that ${\cal A}^{\rm s}$  describes the coherent transfer of a Cooper pair from the right to the left lead, while ${\cal A}^{\rm d}$ describes first a coherent Cooper pair transfer from the right lead to the dot, where coherence is lost, then a second coherent transfer from the dot to the left lead. Both contributions to  the Josephson current, $J^{\rm s}$ and $J^{\rm d}$   [see Eqs. (\ref{Jtot}) and (\ref{Jsd})],  are suppressed to a certain extent by the spin-precession (as compared with their respective values in the absence of the SOI).  Which of them is more severely affected is determined by the {\it geometry} of the junction and the symmetry direction of the SOI. This feature is discussed in Sec. \ref{Results}.

The effect of the spin-orbit coupling on the supercurrent is embedded in the precession of the spins it induces. It is therefore natural
to express the current, in particular the precession factors,  in terms of orientations of the effective magnetic fields  responsible for the precession in the left and right SOI-active nanowires. Indeed,  upon carrying out the spin summations in Eqs. (\ref{Asm}) and (\ref{Adm}) {\it within each reservoir} [i.e., on $\sigma_{L}$ and $\sigma_{R}$ using Eq. (\ref{W}) for ${\bf W}$], one finds that these factors can be expressed in terms of the vectors ${\bf V}_{L}$ and ${\bf V}_{R}$, that represent the effective magnetic fields, 
\begin{align}
&\sum_{\sigma^{}_{L}}{\rm sgn}(\sigma^{}_{L})
 [({\bf W}^{LD}_{})^{\ast}_{}]^{}_{\sigma^{}_{L}\sigma'}
[{\bf W}^{LD}_{}]^{}_{\sigma^{}_{L}\sigma}\equiv[\sig\cdot{\bf V}^{}_{L}]^{}_{\sigma'\sigma}\ ,\nonumber\\
&\sum_{\sigma^{}_{R}}{\rm sgn}(\sigma^{}_{R})
 [{\bf W}^{DR}_{}]^{}_{\sigma\sigma^{}_{R}}
[({\bf W}^{DR}_{})^{\ast}_{}]^{}_{\sigma'\sigma^{}_{R}}\equiv[\sig\cdot{\bf V}^{}_{R}]^{}_{\sigma\sigma'}\ \ .
\label{VLR}
\end{align}
The vectors ${\bf V}_{L(R)}$ are determined by the detailed form of the SOI, Eq. (\ref{W}), 
\begin{align}
{\bf V}^{}_{L(R)}&=2b^{}_{L(R)z}{\bf b}^{}_{L(R)}\nonumber\\
&+2a^{}_{L(R)}{\bf b}^{}_{L(R)}\times\hat{\bf z}+\hat{\bf z}(1-2b^{2}_{L(R)})\ .
\label{Vlr}
\end{align}
Inserting 
Eqs. (\ref{VLR})
into  the spin-precession factors Eqs. (\ref{Asm}) and (\ref{Adm}), one finds
\begin{align}
{\cal A}^{\rm s}_{}&={\bf V}^{}_{L}\cdot{\bf V}^{}_{R}\ ,\nonumber\\
{\cal A}^{\rm d}_{}&=V^{}_{Lz}V^{}_{Rz}\ .
\label{AVV}
\end{align}
It is thus seen that due to the Pauli exclusion principle, only the components of the vectors ${\bf V}_{L}$ and ${\bf V}_{R}$ that are parallel to the quantization axis on the dot contribute to the spin-precession factor arising from the tunneling process in which the dot is doubly occupied, whereas all components of these vectors participate in the spin-precession factor of the sequential transmission. We also note that the difference between the two spin-precession factors disappears when either of the vectors ${\bf V}_{L(R)}$ or both are directed along $\hat{\bf z}-$axis (which is the situation in the absence of the SOI coupling).

\section{Results}
\label{Results}



Although it is possible in principle to calculate an effective SOI  {\it ab initio},  it is rather ubiquitous  to adopt the phenomenological Hamiltonian  proposed in 
Ref.~\onlinecite{Bychkov1984}.   This Hamiltonian (named after Rashba) is valid for systems with a single high-symmetry axis that lack spatial inversion symmetry. For  an electron of an effective mass $m^{\ast}$ and  momentum ${\bf p}$ propagating along  a wire where the SOI is active, it reads
\begin{align}
{\cal H}_{\rm so}=\frac{ k^{\rm Rashba}_{\rm so}}{m^{\ast}}\sig\cdot({\bf p}\times\hat{\bf n})\ .
\label{Hso}
\end{align}
Here   $\hat{\bf n}$ is a unit vector along the symmetry axis (the $\hat{\bf c}-$axis in  hexagonal wurtzite crystals, the growth direction in a semiconductor heterostructure, the direction of an external electric field),  and $k^{\rm Rashba}_{\rm so}$ is the strength of the SOI  in units of inverse length, which is usually taken from experiments. [\onlinecite{Sato2001,Beukman2017}]
By exploiting  the Hamiltonian (\ref{Hso}) to find the propagator along a one-dimensional wire, one obtains that the tunneling amplitude is [\onlinecite{Shahbazyan1994,Shekhter2013}]
\begin{align}
{\bf t}^{}_{\bf k}= it^{}_{L}e^{ik^{}_{\rm F}r^{}_{L}}\exp[ik^{\rm Rashba}_{\rm so}{\bf r}^{}_{L}\times\hat{\bf n}\cdot\sig]\ ,
\label{GFL}
\end{align}
with an analogous form for ${\bf t}_{\bf p}$. Here ${\bf r}_{L(R)} $ is the radius vector pointing from the dot to the left (right) reservoir  along the wire, and $k_{\rm F}$ is the Fermi wave vector of an electron traversing this nanowire. The linear Dresselhaus SOI [\onlinecite{Dresselhaus1955}] gives rise to a comparable form for ${\cal H}_{\rm so}$ [\onlinecite{Ora2005}].
Another source for SOI's are strains, created for instance when a single flat graphene ribbon is rolled up to form a tube.
This type of SOI was modeled by the Hamiltonian
\begin{align}
{\cal H}^{\rm strain}_{\rm so}=
\frac{k^{}_{\rm F} k^{\rm strain}_{\rm so}}{m^{\ast}}\sig\cdot\hat{\bf n}\ ,
\label{tube}
\end{align}
 where
$k^{\rm strain}_{\rm so}$ is a phenomenological parameter that gives the strength of the SOI in units of inverse length
and the unit vector $\hat{\bf n}$ points along the nanotube [\onlinecite{Rudner2010,Flensberg2010}].

In the case of the Rashba SOI, the spin-dependent factor in the tunneling amplitude Eq.  (\ref{GFL}) can be written in the form
\begin{align}
{\bf W}^{}_{}&=\exp[i\tilde{\alpha}^{}_{}\sig\cdot\hat{\bf v}^{}_{}]
=\cos(\tilde{\alpha})+i\sin(\tilde{\alpha})\sig\cdot\hat{\bf v}^{}_{}\ ,
\label{genso}
\end{align}
where  $\tilde{\alpha}$ is proportional to the strength of the spin-orbit coupling and both $\tilde{\alpha}$  and the unit vector $\hat{\bf v}$ are  determined by the symmetry direction of the SOI and the geometrical configuration of the junction: $\tilde{\alpha}=k^{\rm Rashba}_{\rm so}d\sin\phi_{}$, where $d_{}$ is the length of the SOI-active bridge, and $\phi_{}$ is the angle between the wire direction and $\hat{\bf n}$, the symmetry axis of the interaction. 

The parameters that characterize the spin-orbit coupling  can be controlled experimentally. The capability to tune the strength of the spin-orbit interaction electrostatically was first demonstrated in Ref.~\onlinecite{Nitta1997} on
inverted In$^{}_{0.53}$Ga$^{}_{0.47}$As/In$^{}_{0.52}$Al$^{}_{0.48}$As heterostructures.
A more recent experimental evidence is found in  Ref.~\onlinecite{Sato2001}, that describes
measurements on the inversion layer of a In$^{}_{0.75}$Ga$^{}_{0.25}$As/In$^{}_{0.75}$Al$^{}_{0.25}$As semiconductor heterostructure, and  in  
Ref.~\onlinecite{Beukman2017},  which reports on a dual-gated InAs/GaSb quantum well.
The electrodes in these experiments 
are two-dimensional electron  (or hole)
gas bulk conductors; two gate electrodes are used to tune both the carriers' concentration and the strength of the SOI. The  spin-orbit coupling is  characterized by the ``Rashba parameter" $\alpha^{}_{\rm R}$, whose relation to $k^{\rm Rashba}_{\rm so}$ is
\begin{align}
k^{\rm Rashba}_{\rm so}=m^{\ast}\alpha^{}_{R}/\hbar^{2}\ ,
\label{Rpara}
\end{align} 
where $\alpha^{}_{\rm R}$ is measured in [meV \AA] (we keep $\hbar$ in the  experimental estimations).
In Ref.~\onlinecite{Sato2001}, the spin-orbit coupling constant $\alpha$ varied with the gate voltage between roughly 150 and 300 
meV \AA.
Attributing $\alpha$ mainly to the Rashba SOI parameter $\alpha^{}_{R}$, and using $m^{\ast}=0.04 \ m$
($m$ is the free-electron mass),
one concludes that if a weak link were to be electrostatically defined 
in such a system, then $k_{\rm so}^{\rm Rashba}d$ could be varied from $\sim 8$ to $\sim 16$ for $d=1\ \mu m$. 
The ratio of the effective mass to the free-electron mass quoted in Ref.~\onlinecite{Beukman2017} is comparable;
in this quantum well the Dresselhous SOI was kept constant 
while $\alpha^{}_{R}$ could be varied between 53 and 75 
meV \AA, leading to 
$k_{\rm so}^{\rm Rashba}d$ that
varies from $\sim 4$ to $\sim 8$ for $d=1\ \mu m$.

As an explicit example, we consider a weak link of the form of two straight segments, whose 
SOI's parameters are   $\tilde{\alpha}_{L(R)}$ and $\hat{\bf v}_{L(R)}$. One then finds that the two spin-precession factors, Eqs. (\ref{AVV}),  are
[see Eqs. (\ref{Vlr}) and (\ref{genso})]
\begin{align}
{\cal A}^{\rm d}_{}=&[\cos(2\tilde{\alpha}^{}_{L})+2v'^{2}_{Lz}
][\cos(2\tilde{\alpha}^{}_{R})+2v'^{2}_{Rz}] \ , 
\label{Adv}
\end{align}
and
\begin{align}
&{\cal A}^{\rm s}_{}={\cal A}^{\rm d}_{}
+4\sin(\tilde{\alpha}^{}_{L})\sin(\tilde{\alpha}^{}_{R})\nonumber\\
&\times\{
[\hat{\bf v}^{}_{L}\times\hat{\bf v}^{}_{R}]^{}_{z}[\cos(\tilde{\alpha}^{}_{R})v'^{}_{Lz}-\cos(\tilde{\alpha}^{}_{L})v'^{}_{Rz}]\nonumber\\
&+{\bf v}^{}_{L\perp}\cdot{\bf v}^{}_{R\perp}[\cos(\tilde{\alpha}^{}_{L})\cos(\tilde{\alpha}^{}_{R})+v'^{}_{Lz}v'^{}_{Rz}]\}\ ,
\label{Asv}
\end{align}
where
\begin{align}
v'^{}_{L(R)z}=v^{}_{L(R)z}\sin(\tilde{\alpha}^{}_{L(R)})\ .
\label{vz}
\end{align}
(The notation $\perp$ indicates the part of the vector normal to $\hat{\bf z}$.)
For instance, when both $\hat{\bf v}^{}_{L}$ and $\hat{\bf v}^{}_{R}$
are along $\hat{\bf z}$, then ${\cal A}^{\rm s}={\cal A}^{\rm d}=1$, and the effect of the SOI disappears.
On the other hand, when the spin-orbit coupling is due to the Rashba interaction with an electric field directed along $\hat{\bf z}$ and the junction is lying in the XY plane (see Fig.~\ref{Fig1}), then $\hat{\bf v}_{L}$ and $\hat{\bf v}_{R}$ are in the XY plane. In that case the spin-precession factors are
\begin{align}
{\cal A}^{\rm d}_{}&=\cos(2\tilde{\alpha}^{}_{L})\cos(2\tilde{\alpha}^{}_{R})\ ,\nonumber\\
{\cal A}^{\rm s}&={\cal A}^{\rm d}_{}+\sin(2\tilde{\alpha}^{}_{L})\sin(2\tilde{\alpha}^{}_{R})\cos(\theta)\ ,
\label{finalasad}
\end{align}
where $\theta$ is the angle between $\hat{\bf v}_{L}$ and $\hat{\bf v}_{R}$.
We plot in Fig.~\ref{FigJ2}
the normalized Josephson current, Eq.  (\ref{NorJ}), for $\tilde{\alpha}_{R}=\tilde{\alpha}_{L}=\tilde{\alpha}$ and the spin-precession factors as given in Eqs. (\ref{finalasad}),  as a function of the angle $\theta$, for various values of   the spin-orbit strength $\tilde{\alpha}$.  One notes the  change of sign of the normalized Josephson current, as a function of the angle $\theta$ between $\hat{\bf v}_{L}$ and $\hat{\bf v}_{R}$.

\begin{figure}[htp]
\includegraphics[width=3.in]{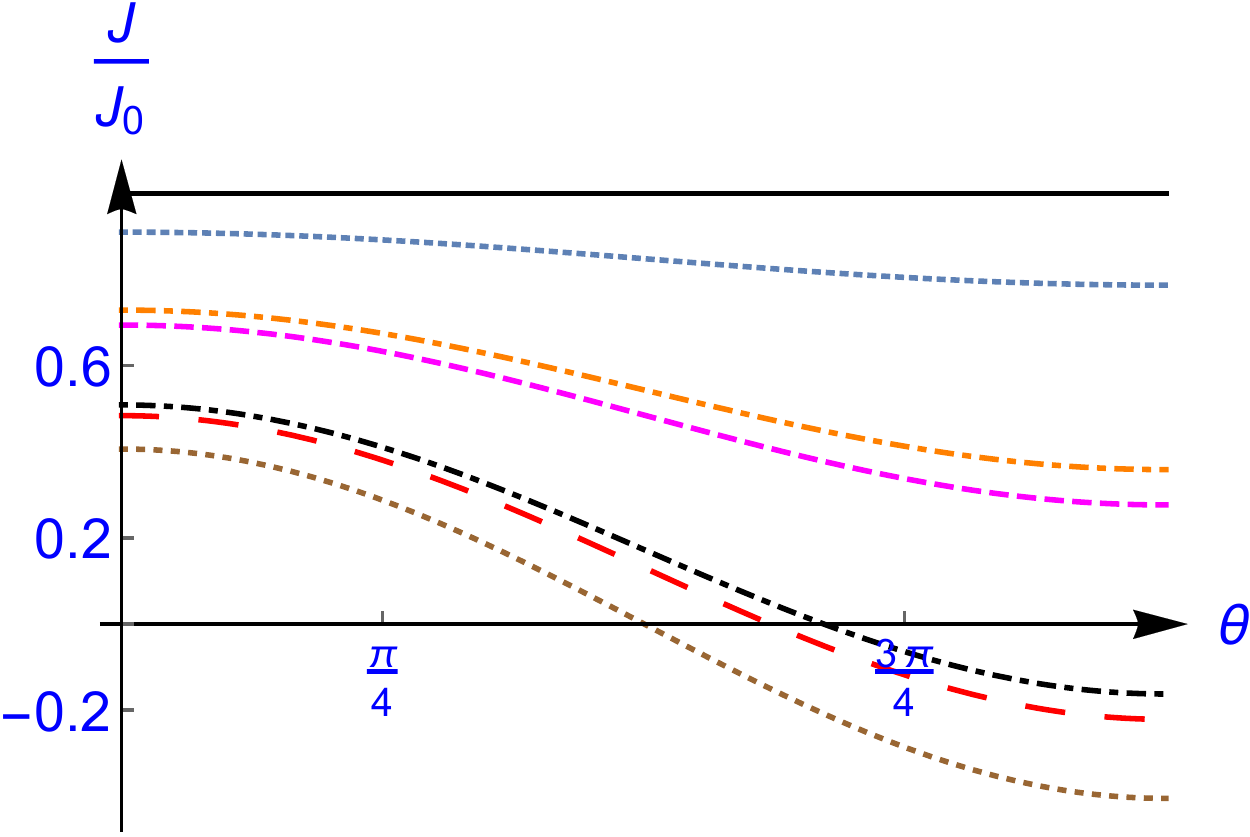}
\caption{(Color online) The normalized Josephson current, Eq. (\ref{NorJ}), as a function of the angle $\theta$ between $\hat{\bf v}_{L}$ and $\hat{\bf v}_{R}$ [see Eqs. (\ref{finalasad})] for various values of $\tilde{\alpha}_{R}=\tilde{\alpha}_{L}=\tilde{\alpha}$. Straight (black) line -- $\tilde{\alpha}=0$, tiny-dashed (blue) line --  $\tilde{\alpha}=0.2$,  medium-dashed (magenta) curve -- 
$\tilde{\alpha}=0.4$, large-dashed (red) curve -- 
$\tilde{\alpha}=0.6$,  dotted (brown) curve -- 
$\tilde{\alpha}=0.8$,    dot-dashed (black) curve -- 
$\tilde{\alpha}=1.$,  dot-dashed (orange) curve -- 
$\tilde{\alpha}=1.2$.      The parameters that determine Eqs. (\ref{Fsm}) and (\ref{Fdm}) are  $\epsilon/\Delta=0$ and $U/\Delta=5$.   }
\label{FigJ2}
\end{figure}

Figure \ref{FigJ3}  displays the normalized Josephson current as a function of {\it both} $\theta$ and the spin-orbit coupling constant $\tilde{\alpha}$. Here 
one notes  the conspicuous 
oscillations as a function of $\tilde{\alpha}$, which arise from the trigonometric functions in Eqs. (\ref{finalasad}).
Figures \ref{FigJ2} and \ref{FigJ3} exemplify the possibility to vary  the supercurrent in our device  mechanically,  by bending the bridge (and thus changing $\theta$) connecting the superconductors.

\begin{figure}[htp]
\includegraphics[width=3.4in]{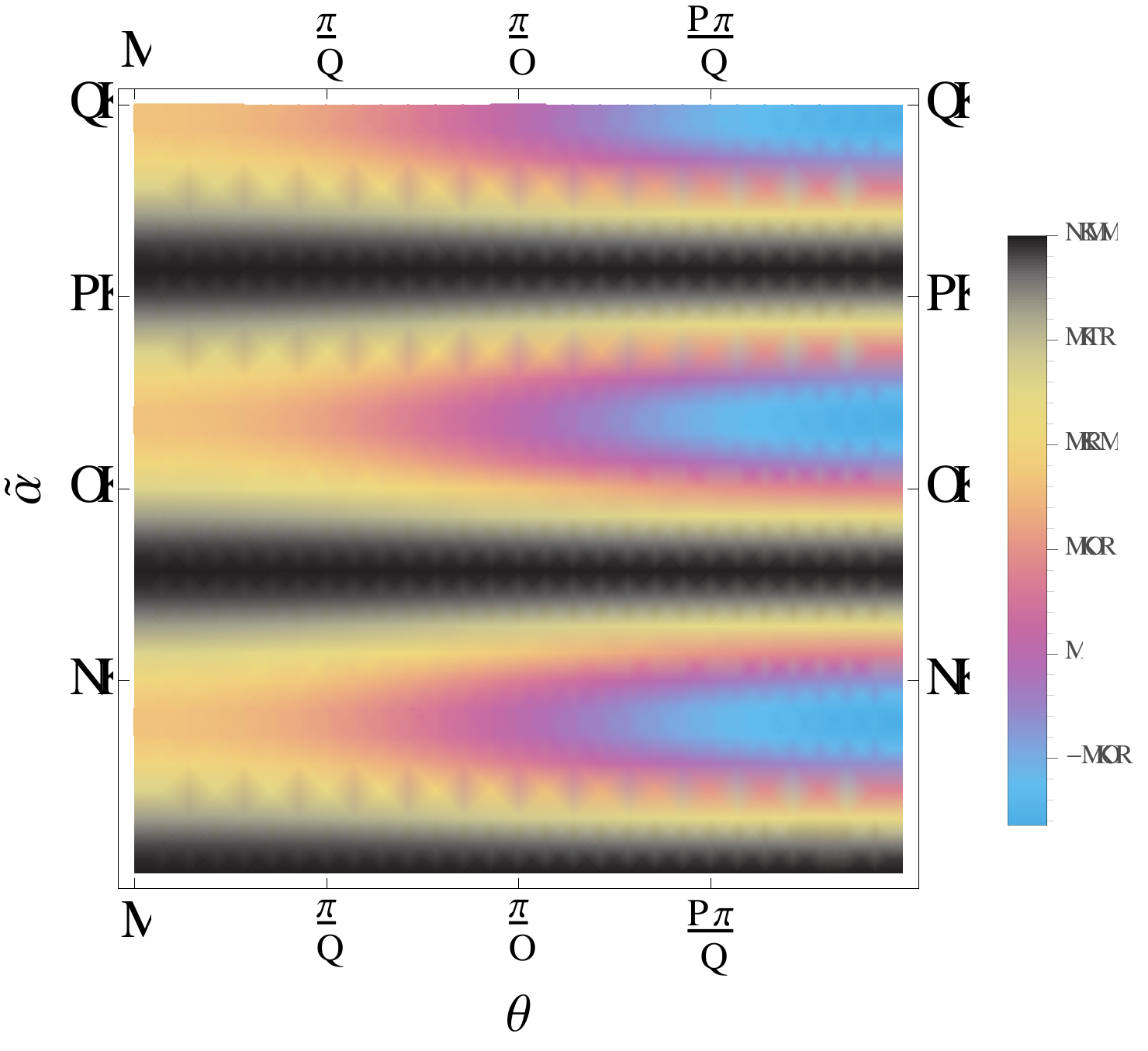}
\caption{(Color online) A density plot of the normalized Josephson current, Eq. (\ref{NorJ}), as a function of the angle $\theta$ between $\hat{\bf v}_{L}$ and $\hat{\bf v}_{R}$ and the spin-orbit coupling constant, $\tilde{\alpha}$ [see Eqs. (\ref{finalasad})]. The parameters that determine Eqs. (\ref{Fsm}) and (\ref{Fdm}) are  $\epsilon/\Delta=0$ and $U/\Delta=5$.}
\label{FigJ3}
\end{figure}

As mentioned, the coupling constant of the SOI can be manipulated experimentally by varying gate voltages. Additional functionality is obtained  when the {\it orientations} of these electric fields (induced by the gates)  in the two arms of the bridge are made to be different. The simplest example is when the two arms of the bridge lie along the $\hat{\bf x}-$axis, with the electric field on the left nanowire directed along $\hat{\bf z}$, while that on the right one is in the $YZ-$plane, making an angle $\gamma$ with the electric field on the left wire. 
In this configuration, $\hat{\bf v}_{L}=-\hat{\bf y}$ and $\hat{\bf v}_{R}=-\hat{\bf y}\cos(\gamma)+\hat{\bf z}\sin(\gamma)$.   Using these expressions in Eqs. (\ref{Adv}), (\ref{Asv}), and (\ref{vz})
gives
\begin{align}
{\cal A}^{\rm d}_{}&=\cos(2\tilde{\alpha})[\cos(2\tilde{\alpha})+2\sin^{2}_{}(\tilde{\alpha})
\sin^{2}(\gamma)]\ ,\nonumber\\
{\cal A}^{\rm s}_{}&=
{\cal A}^{\rm d}_{}+\sin^{2}_{}(2\tilde{\alpha})\cos(\gamma)\ , 
\label{ele}
\end{align}
where for simplicity we have assumed that 
$\tilde{\alpha}_{L}=k^{\rm Rashba}_{\rm so}d^{}_{L}$ and $\tilde{\alpha}_{R}=k^{\rm Rashba}_{\rm so}d^{}_{R}$ coincide, i.e.,  $\tilde{\alpha}_{L}=\tilde{\alpha}_{R}\equiv\tilde{\alpha}$.
The normalized Josephson current pertaining to this configuration, as a function of the angle $\gamma$ between the electric fields, is shown in Fig.~\ref{FigJ4} for various values of $\tilde{\alpha}$. Its oscillation with respect to both $\gamma$ and $\tilde{\alpha}$ is displayed in Fig.~\ref{FigJ5}. Importantly,   all our  illustrations are based on  gate-controlled SOI strengths that are amenable in experiment.

\begin{figure}[htp]
\includegraphics[width=3.in]{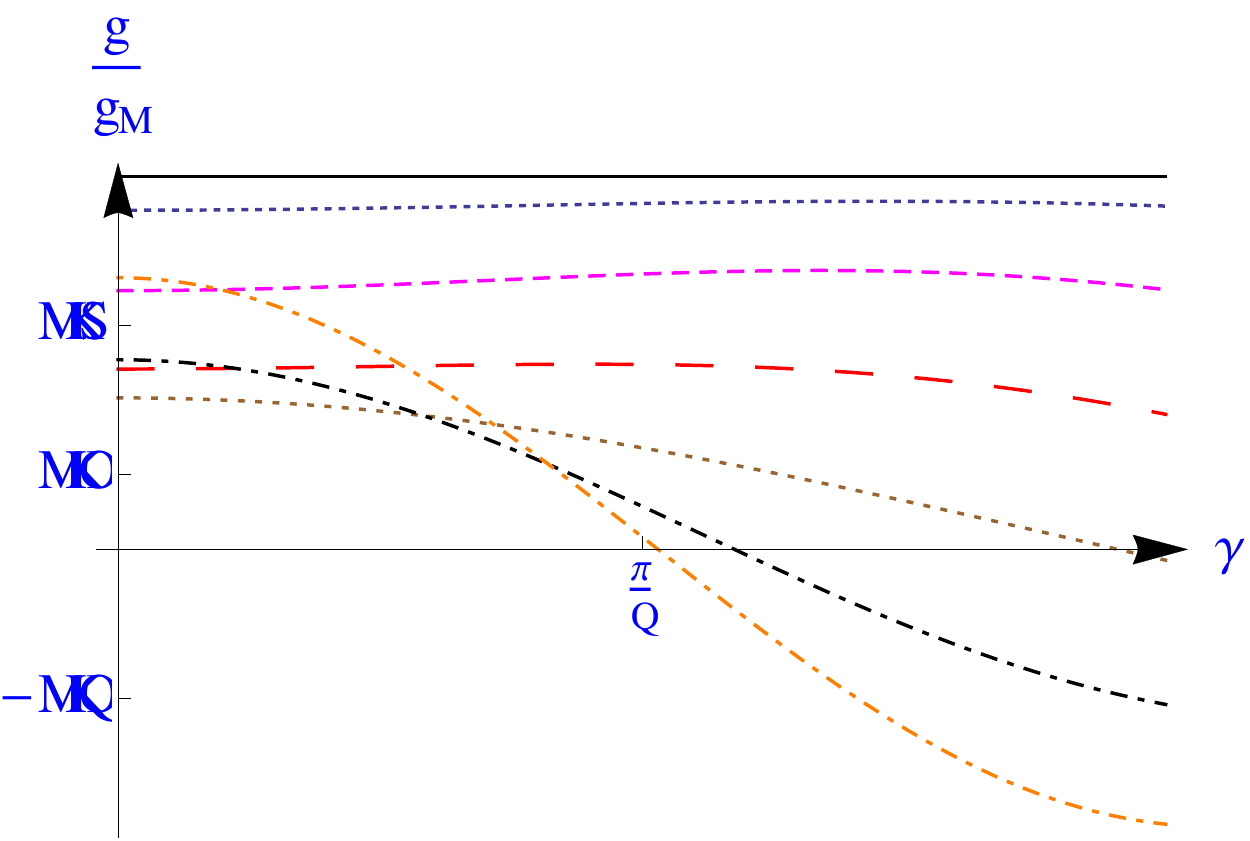}
\caption{(Color online) The normalized Josephson current, Eq. (\ref{NorJ}), as a function of the angle $\gamma$, the `mis-orientation' of the electric fields on the left and right nanowires [see Fig.~\ref{Fig1} and   Eqs.~(\ref{ele})] for various values of $\tilde{\alpha}_{R}=\tilde{\alpha}_{L}=\tilde{\alpha}$. Straight (black) line -- $\tilde{\alpha}=0$, tiny-dashed (blue) line --  $\tilde{\alpha}=0.2$,  medium-dashed (magenta) curve -- 
$\tilde{\alpha}=0.4$, large-dashed (red) curve -- 
$\tilde{\alpha}=0.6$,  dotted (brown) curve -- 
$\tilde{\alpha}=0.8$,    dot-dashed (black) curve -- 
$\tilde{\alpha}=1.$,  dot-dashed (orange) curve -- 
$\tilde{\alpha}=1.2$.      The parameters that determine Eqs. (\ref{Fsm}) and (\ref{Fdm}) are  $\epsilon/\Delta=0$ and $U/\Delta=5$.   }
\label{FigJ4}
\end{figure}

\begin{figure}[htp]
\includegraphics[width=3.4in]{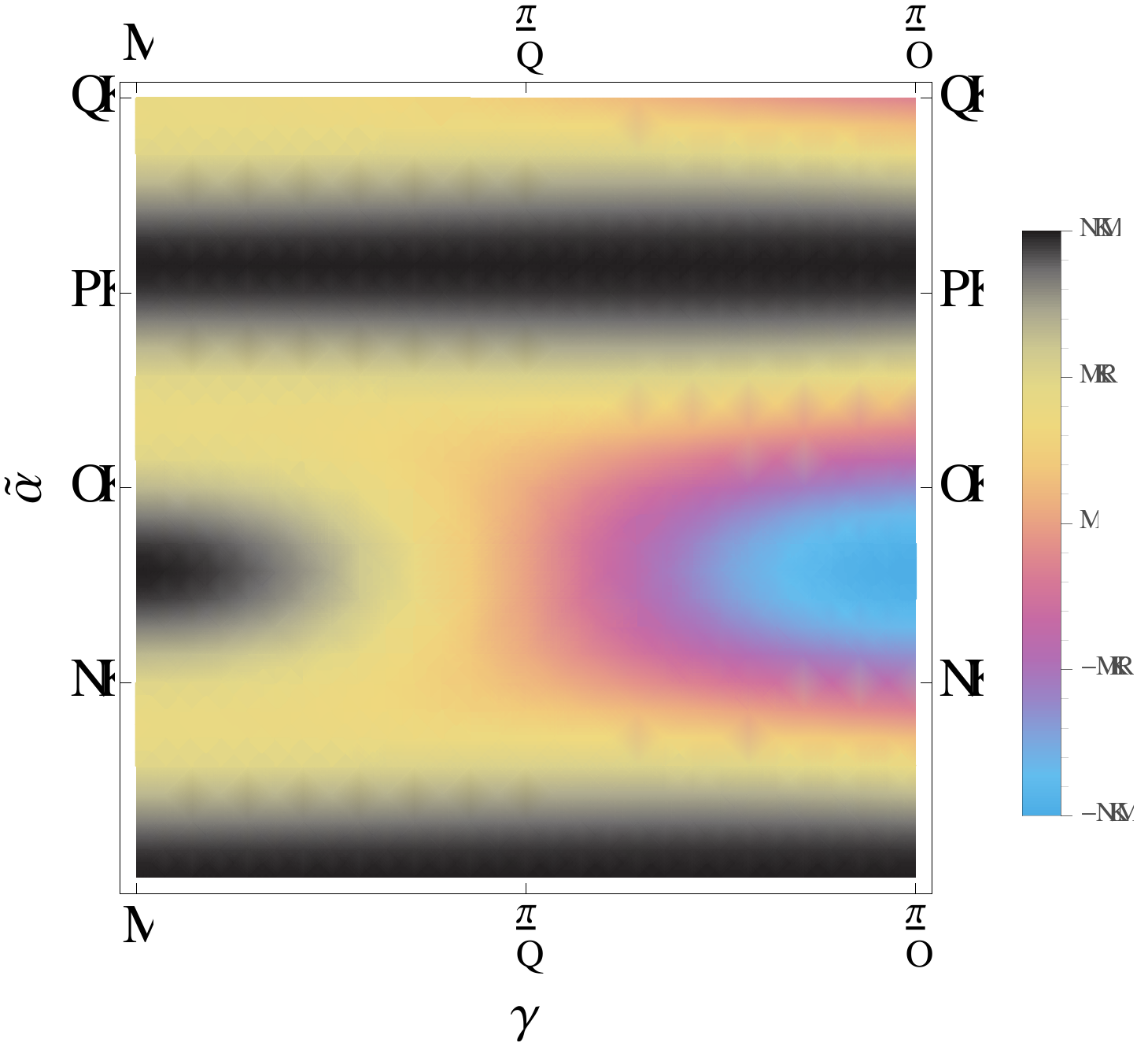}
\caption{(Color online) A density plot of the normalized Josephson current, Eq. (\ref{NorJ}), as a function of the angle $\gamma$ between indicating the `mis-orientation' of the electric fields on the two nanowires,  and the spin-orbit coupling constant, $\tilde{\alpha}$ [see Eqs. (\ref{ele})]. The parameters that determine Eqs. (\ref{Fsm}) and (\ref{Fdm}) are  $\epsilon/\Delta=0$ and $U/\Delta=5$.}
\label{FigJ5}
\end{figure}

\section{Discussion}
\label{Discussion}

We have considered the spin splitting of Cooper pairs that carry a supercurrent through a weak-link Josephson junction. Our main result, illustrated in Figs. \ref{FigJ2} -- \ref{FigJ5}, is the rich oscillatory dependence of the normalized Josephson current, $J/J_{0}$ [see Eq. (\ref{NorJ})], on both the spin-orbit coupling constant $\tilde{\alpha}$ and the geometrical properties of the junction. In the example illustrated in Fig.~\ref{Fig1}, the latter variation is manifested in the dependence of $J/J_{0}$ on the bending angle  $\theta$ between $\hat{\bf v}_{L}$ and  $\hat{\bf v}_{R}$, which are normal to the wires connecting  the dot with the left and right reservoirs, respectively,  and are  lying in the the plane of the junction.
As seen in Fig.~\ref{FigJ2}, 
for certain specific values of $\theta$ and the spin-orbit coupling strength $\tilde{\alpha}$ the current vanishes. Another possibility to manipulate the geometry is to `mis-orient' the electric fields that give rise to the spin-orbit interactions on the weak link. Figures \ref{FigJ4} and \ref{FigJ5} display the dependence of the supercurrent on the angle in-between these two fields.

The oscillatory dependence of the supercurrent on the SOI strength (i.e., the dependence on $\tilde{\alpha}$ in Figs. \ref{FigJ2} -- \ref{FigJ5})  results from a rather complex interference between different transmission events: the single-electron transmission one, that yields $J_{}^{\rm s}$ , and the double-electron transmission that gives $J_{}^{\rm d}$, Eqs. (\ref{Jsd}). In the single-electron transmission channel the two electrons are transferred sequentially one by one, so that at any time during the tunneling there is only one electron on the bridge. By contrast, in the other transmission channel both electrons appear in the link for some period of time, which means that in the Coulomb-blockade limit the transfer of Cooper pairs in this channel is completely suppressed. 
As the Coulomb blockade is lifted, the probability of pairs to be transferred in the double-electron tunneling increases. As seen from 
Eqs. (\ref{Asm1}), (\ref{Alr}),  and (\ref{adm}), the Pauli principle operating on the dot breaks the coherence of the pair transfer in the double-electron tunneling process, but does not ruin completely the contribution to the Josephson current.

The pronounced oscillations of the supercurrent and the sign reversal can be observed for plausible lengths of the weak link, of the order of a micron, supposedly achievable by suitably-designed geometries of the gates. 
The magnitude of the Josephson current through a quantum dot is set by the functions $F^{\rm s}$ and $F^{\rm d}$, Eqs. (\ref{Fsm}) and (\ref{Fdm}),  that are derived for short weak links [\onlinecite{Glazman1989}]. However, whereas the restriction on the length $d$ of the bridge might be strict, $d\ll\xi$ for the orbital part ($\xi$ is the superconducting coherence length), it is far weaker for the spin-dependent part: $k^{\rm Rashba}_{\rm so}d\ll k^{}_{\rm F}\xi$,  since the spin-precession factors ${\cal A}^{\rm s}$ and ${\cal A}^{\rm d}$  are not sensitive to the energy dependence of the transmission amplitude [\onlinecite{Shekhter2017}.
Our results indicate interesting phenomena caused by SOI-induced spin polarization of Cooper pairs.

An intriguing feature of our result concerns the spin-polarization created on the dot due to the superconducting proximity effect in conjunction with the spin-orbit coupling. Calculating this polarization may require higher-orders in the tunneling, which are beyond the scope of the present analysis.

\begin{acknowledgments}

We thank the Computational Science Research Center in Beijing  for the hospitality that allowed for the 
accomplishment of this project. RIS and MJ thank the IBS Center for Theoretical Physics of Complex Systems, 
Daejeon, Rep. of Korea, and OEW and AA thank the Dept. of Physics, Univ. of Gothenburg, for hospitality. 
This work was partially supported by the Swedish Research
Council (VR), by the Israel Science Foundation
(ISF), by the infrastructure program of Israel's
Ministry of Science and Technology under contract
3-11173, by the Pazi Foundation, 
and by the Institute for Basic Science, Rep. of Korea (IBS-R024-D1).

\end{acknowledgments}

 \appendix


\section{Time-reversal symmetry and the tunneling amplitudes}
\label{TRtun}

Here we discuss the effect of the time-reversal transformation on  the tunneling amplitudes of Eq. (\ref{Htun}) as given in Eqs. (\ref{GF}) and (\ref{W}), and prove Eq. (\ref{IMR}). Consider for instance 
$[{\bf t}^{LD}_{\bf k}]_{\sigma\sigma'}$,  the probability amplitude
for an electron to go from the the state $|\sigma'\rangle$ on the dot  
to the state $|{\bf k},\sigma\rangle$  on the left lead. 
We denote by an overline the quantities related to the time-reversed process. Thus, 
$[\overline{\bf t}^{LD}_{\bf k}]_{\overline{\sigma}\overline{\sigma}'}$ is
the probability amplitude for 
the time-reversed process which takes 
an electron from 
the time-reversed state of $|\sigma'\rangle$ on the dot,  -- i.e., from $|\overline{\sigma'}\rangle$ -- 
to the time-reversed state of
$|{\bf k},\sigma\rangle$
in the left  lead, that is,   to
$
|-{\bf k},\overline{\sigma}\rangle$. The time-reversal transformation is given in Eq. (\ref{tbar}), and is reproduced here for clarity, 
\begin{align}
\overline{\bf t}^{LD}_{\bf k}=\hat{\bf T}
{\bf t}^{LD}_{\bf k}\hat{\bf T}^{-1}_{}\ ,
\end{align}
where $\hat{\bf T}=K(i\sigma^{}_{y})$ is the time-reversal operator; $K$ is the complex conjugation operator,  and $\sigma_{y}$ is the Pauli matrix. Hence,  
\begin{align}
|\overline{\up}\rangle\equiv(i\sigma^{}_{y})|\up\rangle
& 
=-|\down\rangle\ ,\nonumber\\
|\overline{\down}\rangle\equiv(i\sigma^{}_{y})|\down\rangle&
=|\up\rangle\ .
\label{upb}
\end{align}
The spin-orbit interaction by itself is time-reversal symmetric, i.e., 
its matrix part ${\bf W}$ [see Eqs. (\ref{GF}) and (\ref{W})] is invariant under  the time-reversal transformation, $\overline{\bf W}={\bf W}$, while the scalar factor (i.e., 
$it_{L(R)}\exp[-ik_{\rm F}d_{L(R)}]$ is complex-conjugated.
It remains to find the tunneling amplitude in the basis of the time-reversed states.
To this end we use the generic form of the linear SOI, ${\bf W}$ [see Eq. (\ref{W})].
Using Eqs. (\ref{upb}), one finds
\begin{align}
[\overline{\bf W}^{L(R)D}_{}]^{}_{\overline{\sigma}\overline{\sigma}'}
=[({\bf W}^{L(R)D}_{})^{\ast}_{}]^{}_{\sigma\sigma'}
\ ,
\end{align}
which leads to the relation Eq. (\ref{IMR}).

\begin{widetext}
\section{Expansion of the particle current}
\label{Pert}

Upon using the expansion Eq. (\ref{S}) in the expression (\ref{JL}) for the particle current, one finds quite a number of terms. However, only four of them describe the transfer of Cooper pairs at thermal equilibrium, 
\begin{align}
&i\int_{-\infty}^{t}dt^{}_{1}\int_{-\infty}^{t^{}_{1}}dt^{}_{2}\int_{-\infty}^{t^{}_{2}}dt^{}_{3}
[\langle{\cal H}^{}_{LD}(t){\cal H}^{}_{DR}(t^{}_{1}){\cal H}^{}_{LD}(t^{}_{2}){\cal H}^{}_{DR}(t^{}_{3})
\rangle +\langle{\cal H}^{}_{LD}(t){\cal H}^{}_{LD}(t^{}_{1}){\cal H}^{}_{DR}(t^{}_{2}){\cal H}^{}_{DR}(t^{}_{3})
\rangle]\nonumber\\
&-i\int_{-\infty}^{t}dt^{}_{1}\int_{-\infty}^{t}dt^{}_{2}\int_{-\infty}^{t^{}_{2}}dt^{}_{3}
\langle{\cal H}^{}_{LD}(t^{}_{1}){\cal H}^{}_{LD}(t){\cal H}^{}_{DR}(t^{}_{2}){\cal H}^{}_{DR}(t^{}_{3})
\rangle
\nonumber\\
&+
i\int_{-\infty}^{t}dt^{}_{1}\int_{-\infty}^{t^{}_{1}}dt^{}_{2}\int_{-\infty}^{t}dt^{}_{3}
\langle{\cal H}^{}_{LD}(t^{}_{2}){\cal H}^{}_{DR}(t){\cal H}^{}_{LD}(t){\cal H}^{}_{DR}(t^{}_{3})
\rangle
\ .
\label{detex}
\end{align}
[Recall that the dot is empty in the decoupled state of the junction [\onlinecite{occupied.dot}] Examining the expressions in Eq. (\ref{detex}) in conjunction with Eq. (\ref{Htun}) shows the following features. (i) Each of the terms corresponds to the annihilation of a pair of electrons in the right reservoir and the creation of a pair in the left reservoir. [Note that the particle current, Eq. (\ref{JL}), requires the imaginary part of (\ref{detex}), which means that it includes also analogous terms corresponding to  a pair creation in the right reservoir, and a pair  annihilation in the left one.] As the electrons in each pair are in two time-reversed states,  the two  tunneling amplitudes  are related according to Eq. (\ref{IMR}). For instance,  the first term in Eq. (\ref{detex}) is
\begin{align}
&i\sum_{{\bf k},{\bf p},\sigma^{}_{L},\sigma^{}_{R}}
\sum_{\sigma,\sigma'}
[\overline{\bf t}^{LD}_{\bf k}]^{}_{\overline{\sigma}^{}_{L}\overline{\sigma}'}
[\overline{\bf t}^{DR}_{\bf p}]^{}_{\overline{\sigma}'\overline{\sigma}^{}_{R}}
[{\bf t}^{LD}_{\bf k}]^{}_{\sigma^{}_{L}\sigma}
[{\bf  t}^{DR}_{\bf p}]^{}_{\sigma'\sigma^{}_{R}}\nonumber\\
&\times\int_{-\infty}^{t}dt^{}_{1}\int_{-\infty}^{t^{}_{1}}dt^{}_{2}\int_{-\infty}^{t^{}_{2}}dt^{}_{3}
\langle
c^{\dagger}_{-{\bf k}\overline{\sigma}^{}_{L}}(t)d^{}_{\overline{\sigma}'}
(t)
d^{\dagger}_{\overline{\sigma}'}(t^{}_{1})c^{}_{-{
\bf p}\overline{\sigma}^{}_{R}}(t^{}_{1})
c^{\dagger}_{{\bf k}\sigma^{}_{L}}(t^{}_{2})d^{}_{\sigma'}
(t^{}_{2})
d^{\dagger}_{\sigma'}(t^{}_{3})c^{}_{{
\bf p}\sigma^{}_{R}}(t^{}_{3})\rangle\ .
\label{I1s}
\end{align}
(ii) Two of the terms in Eq. (\ref{detex}), the first and the fourth, correspond to sequential tunneling,  in which the dot is only singly occupied 
in the intermediate state. In the other two terms, the dot is doubly occupied in the intermediate state, and therefore the evolution of the spin states of the tunneling pair is disrupted. (iii) The quantum averages of the operators of the reservoirs are nonzero only in the superconducting state, i.e., when both leads are superconducting.

The remaining part of the calculation is routine: using the Bogoliubov transformation, one derives the time-dependent quantum average of the operators of the 
reservoirs. Those on the dot are calculated using the Hamiltonian of the decoupled dot [the first two terms on the right hand-side of Eq. (\ref{H0})]. In this way,  the expression in Eq. (\ref{I1s}) becomes
\begin{align}
-2e^{i(\phi^{}_{R}-\phi^{}_{L})}\sum_{{\bf k},{\bf p}}
|t^{}_{\bf k}|^{2}|t^{}_{\bf p}|^{2}
\frac{\Delta^{}_{L}}{2E{}_{k}}
\frac{\Delta^{}_{R}}{2E{}_{p}}
\frac{1}{E^{}_{k}+\epsilon}
\frac{1}{E^{}_{k}+E^{}_{p}}
\frac{1}{E^{}_{p}+\epsilon}{\cal A}^{\rm s}_{}\ .
\label{ex1}
\end{align}
where $E^{2}_{k(p)}=\epsilon^{2}_{k(p)}+\Delta^{2}_{L(R)}$. For simplicity,  the temperature is set to  zero.  
In deriving this expression,  we have made use of Eq.  (\ref{IMR}), that relates the tunneling amplitudes of two time-reversed events.
The factor ${\cal A}^{\rm s}$ describes the spin precession in the sequential tunneling processes [i.e., the first and the fourth terms in 
Eq. (\ref{I1s})].
Explicitly, 
\begin{align}
{\cal A}^{\rm s}_{}=\frac{1}{2}\sum_{\sigma,\sigma'}\sum_{\sigma^{}_{L},\sigma^{}_{R}}{\rm sgn}(\sigma^{}_{L}){\rm sgn}(\sigma^{}_{R})[({\bf W}^{LD}_{})^{\ast}_{}]^{}_{\sigma^{}_{L}\sigma'}
[{\bf W}^{LD}_{}]^{}_{\sigma^{}_{L}\sigma}
[({\bf W}^{DR}_{})^{\ast}_{}]^{}_{\sigma'\sigma^{}_{R}}
[{\bf W}^{DR}_{}]^{}_{\sigma\sigma^{}_{R}}\ .
\label{As}
\end{align}
As in the absence of the SOI the matrices ${\bf W}$  are all just the unit matrix, the spin-precession factor ${\cal A}^{
\rm s}$ becomes then 1.
The sums over ${\bf k}$ and ${\bf p}$ are carried out assuming that $|t_{{\bf k}({\bf p})}|^{2}$ and the single-particle density of states of the leads can be approximated by their respective values at the Fermi energy, $|t_{L(R)}|^{2}$ and ${\cal N}
_{L(R)}$.  For a short weak link   with $\Delta_{L}=\Delta_{R}\equiv\Delta$ [\onlinecite{Glazman1989}],  these sums then give
$(\Gamma_{L}\Gamma_{R}/\Delta) F^{\rm s}(\epsilon/\Delta)$, where  $\Gamma_{L(R)}=\pi{\cal N}_{L(R)}|t^{}_{L(R)}|^{2}$, and the function $F^{\rm s}$ is
\begin{align}
F^{\rm s}_{}(\widetilde{\epsilon})=
\int_{-\infty}^{\infty}\frac{d\zeta^{}_{k}}{\pi}
\int_{-\infty}^{\infty}\frac{d\zeta^{}_{p}}{\pi}
[({\rm cosh}\zeta^{}_{k}+\widetilde{\epsilon})
({\rm cosh}\zeta^{}_{k}+{\rm cosh}\zeta^{}_{p})
({\rm cosh}\zeta^{}_{p}+\widetilde{\epsilon})]^{-1}\ ,\ \ \widetilde{\epsilon}=\epsilon/\Delta\ .
\label{Fs}
\end{align}
An identical result is obtained for the fourth term in the expansion (\ref{detex}).
The imaginary part of the expression in Eq. (\ref{ex1}) consists of three factors, 
the Josephson amplitude of  the interface between the two superconductors  (i,e., in the absence of  the resonant level on the dot and the SOI), 
$I_{0}=2\sin(\phi_{R}-\phi_{L})[\Gamma^{}_{L}\Gamma^{}_{R}/\Delta]$, the function $F^{\rm s}$ that conveys  the effect of the localized level on the dot, and the spin-precession factor, ${\cal A}^{\rm s}$. The two latter factors are discussed in Sec. \ref{Results}.

The second term in Eq. (\ref{detex}), which pertains to the situation where during the tunneling process the dot is doubly occupied, reads
\begin{align}
&i\sum_{{\bf k},{\bf p},\sigma^{}_{L},\sigma^{}_{R}}
\sum_{\sigma}
[\overline{\bf t}^{LD}_{\bf k}]^{}_{\overline{\sigma}^{}_{L}\overline{\sigma}}
[\overline{\bf t}^{DR}_{\bf p}]^{}_{\overline{\sigma}\overline{\sigma}^{}_{R}}
[{\bf t}^{LD}_{\bf k}]^{}_{\sigma^{}_{L}\sigma}
[{\bf  t}^{DR}_{\bf p}]^{}_{\sigma\sigma^{}_{R}}\nonumber\\
&\times\int_{-\infty}^{t}dt^{}_{1}\int_{-\infty}^{t^{}_{1}}dt^{}_{2}\int_{-\infty}^{t^{}_{2}}dt^{}_{3}
\langle
c^{\dagger}_{-{\bf k}\overline{\sigma}^{}_{L}}(t)d^{}_{\overline{\sigma}}
(t)
c^{\dagger}_{{\bf k}\sigma^{}_{L}}(t^{}_{1})d^{}_{\sigma}
(t^{}_{1})
d^{\dagger}_{\overline{\sigma}}(t^{}_{2})c^{}_{-{
\bf p}\overline{\sigma}^{}_{R}}(t^{}_{2})
d^{\dagger}_{\sigma}(t^{}_{3})c^{}_{{
\bf p}\sigma^{}_{R}}(t^{}_{3})\rangle\ ,
\label{I1d}
\end{align}
where we have taken into account the Pauli principle, and therefore there are only three summations over the spin indices [{\it c.f.} Eq. (\ref{I1s})].
In this case we obtain
\begin{align}
-4e^{i(\phi^{}_{R}-\phi^{}_{L})}\sum_{{\bf k},{\bf p}}
|t^{}_{\bf k}|^{2}|t^{}_{\bf p}|^{2}
\frac{\Delta^{}_{L}}{2E{}_{k}}
\frac{\Delta^{}_{R}}{2E{}_{p}}
\frac{1}{E^{}_{k}+\epsilon}
\frac{1}{2\epsilon+U}
\frac{1}{E^{}_{p}+\epsilon}{\cal A}^{\rm d}_{}\ ,
\label{ex2}
\end{align}
where ${\cal A}^{\rm d}$ describes the spin precession in the  tunneling processes in which the two electrons reside simultaneously on the dot in the intermediate sate [i.e., the second and the third terms in 
Eq. (\ref{I1s})]. Its explicit form is
\begin{align}
{\cal A}^{\rm d}_{}=\frac{1}{2}\sum_{\sigma}\sum_{\sigma^{}_{L},\sigma^{}_{R}}{\rm sgn}(\sigma^{}_{L}){\rm sgn}(\sigma^{}_{R})
[({\bf W}^{LD}_{})^{\ast}_{}]^{}_{\sigma^{}_{L}\sigma}
[{\bf W}^{LD}_{}]^{}_{\sigma^{}_{L}\sigma}
[({\bf W}^{DR}_{})^{\ast}_{}]^{}_{\sigma\sigma^{}_{R}}
[{\bf W}^{DR}_{}]^{}_{\sigma\sigma^{}_{R}}\ .
\label{Ad}
\end{align}
Similar to  ${\cal A}^{\rm s}$, this factor also  becomes 1 in the absence of the SOI.
The sums over ${\bf k}$ and ${\bf p}$ are carried out as explained above. Because of the double occupancy of the dot, the energy denominators in Eq. (\ref{ex2})  differ from those in Eq. (\ref{ex1}). These summations give rise to another  function,  $F^{\rm d}(\epsilon/\Delta,U/\Delta)$, of the energies on the dot [\onlinecite{Glazman1989}]
\begin{align}
F^{\rm d}_{}(\widetilde{\epsilon},\widetilde{U})=
\int_{-\infty}^{\infty}\frac{d\zeta^{}_{k}}{\pi}
\int_{-\infty}^{\infty}\frac{d\zeta^{}_{p}}{\pi}
[({\rm cosh}\zeta^{}_{k}+\widetilde{\epsilon})
(2\widetilde{\epsilon}+\widetilde{U}) 
({\rm cosh}\zeta^{}_{p}+\widetilde{\epsilon})]^{-1}\ ,\ \ \widetilde{\epsilon}=\epsilon/\Delta\ ,\ \ \widetilde{U}=U/\Delta\ .
\label{Fd}
\end{align}
Since the third term in the expansion (\ref{detex}) turns out to be identical to Eq. (\ref{ex2}), it follows that the contribution from these tunneling processes to the Josephson current is again a product of three factors, $I_{0}$, $F^{\rm d}$, and ${\cal A}^{\rm d}$. 

\end{widetext}



\begin{thebibliography}{99}
\bibitem{axis}
The quantization axis of the spins is taken to be the same, along the $\hat{\bf z}-$direction, in  both the superconducting leads and the dot (it can be set, e.g., by a small external magnetic field along that direction).
\bibitem{occupied.dot}
The other possibility,  of a dot which is initially occupied, is also of interest. 
We hope to pursue the spin-splitting phenomenon for this configuration in the future.
\end{thebibliography}
\end{document}